\documentclass[floatfix,twocolumn,amsmath,amssymb,aps,prl,showpacs,letter,superscriptaddress]{revtex4}
\usepackage{latexsym,epsfig,bm,times,psfrag,subfigure}

\usepackage{color}
\usepackage{ulem}

\newcommand{\tso}{Tb$_{2}$Sn$_{2}$O$_{7}$}

\newcommand{\tto}{Tb$_{2}$Ti$_{2}$O$_{7}$}

\begin{document}

\title{Lack of Evidence for a Singlet Crystal Field Ground State in the
 Tb$_2$Ti$_2$O$_7$ Magnetic Pyrochlore}

\author{B. D. Gaulin}
\affiliation{Department of Physics and Astronomy, McMaster University, Hamilton, ON, L8S 4M1, Canada.}
\affiliation{Brockhouse Institute for Materials Research, McMaster University, Hamilton, ON, L8S 4M1, Canada.}
\affiliation{Canadian Institute for Advanced Research, 180 Dundas Street West, Suite 1400, Toronto, ON, M5G 1Z8, Canada.}
\author{J.S. Gardner}
\affiliation{Indiana University, 2401 Milo B. Sampson Lane, Bloomington, Indiana 47408, USA}
\affiliation{NCNR, NIST, Gaithersburg, Maryland 20899-6102, USA}
\author{P. A. McClarty}
\affiliation{Department of Physics and Astronomy, University of Waterloo, Waterloo, ON, N2L 3G1, Canada.}
\affiliation{Max Planck Institute for the Physics of Complex Systems, N\"{o}thnitzer Str. 38, 01187 Dresden, Germany.}
\author{M. J. P. Gingras}
\affiliation{Canadian Institute for Advanced Research, 180 Dundas Street West, Suite 1400, Toronto, ON, M5G 1Z8, Canada.}
\affiliation{Department of Physics and Astronomy, University of Waterloo, Waterloo, ON, N2L 3G1, Canada.}

\date{\today}

\begin{abstract}
We present new high resolution inelastic neutron scattering data
on the candidate spin liquid Tb$_2$Ti$_2$O$_7$. We find that there is no
evidence for a zero field splitting of the ground state doublet within the $0.2$ K
resolution of the instrument.
This result contrasts with a pair of recent works on
Tb$_2$Ti$_2$O$_7$ claiming that the spin liquid behavior can be attributed to
a $2$ K split singlet-singlet single-ion spectrum at low energies. We also
reconsider the entropy argument presented in Chapuis {\it et al.} as further
evidence of a singlet-singlet crystal field spectrum. We arrive at the
conclusion that estimates of the low temperature residual entropy drawn from
heat capacity measurements are a poor guide to the single ion spectrum without
understanding the nature of the correlations. 
\end{abstract}

\pacs{
75.10.Dg          
75.10.Jm          
75.40.Cx          
 75.40.Gb          
}

\maketitle


{\it Introduction} $-$ In some magnetic systems, the lattice geometry or the competition between different
interactions can dramatically inhibit, or frustrate, the development of long range order.
The failure of a frustrated magnetic system to exhibit magnetic order
down to zero temperature, giving rise to a so-called spin liquid state, 
is one of the most sought after phenomena
among strongly interacting condensed matter systems.
Despite two decades of experimental searches, the number of candidate materials 
that display a spin liquid state remains small  \cite{Balents-Nature}.
The Tb$_2$Ti$_2$O$_7$ insulating material, where magnetic Tb$^{3+}$ ions 
sit on a pyrochlore lattice of corner-sharing tetrahedra, 
is one of these candidates \cite{Gardner-TbTO-PRL-1999}.
Despite a Curie-Weiss temperature, $\theta_{\rm CW} \approx -14$ K set by the magnetic
interactions \cite{Gingras-TbTO-PRB-2000},
Tb$_2$Ti$_2$O$_7$ does not develop long  range order down to at least 50 mK 
\cite{Gardner-TbTO-PRL-1999,Gardner-TbTO-PRB-2003}.
The microscopic mechanism by which Tb$_2$Ti$_2$O$_7$ fails to develop long range
order at a temperature scale of approximately 1 K \cite{Gingras-TbTO-PRB-2000,Hertog-PRL-2000,Kao}
is not understood \cite{TTOHeff}.
Compounding the difficulty in understanding why
Tb$_2$Ti$_2$O$_7$ does not order, one notes that  Tb$_2$Sn$_2$O$_7$,
seemingly closely related at the microscopic level to Tb$_2$Ti$_2$O$_7$, 
develops long range order at 0.87 K \cite{Matsuhira-TbSnO,Mirebeau-TbSnO}.

Since Tb$^{3+}$ is an even electron system (electronic configuration $^7$F$_6$, $L=3$, $S=3$, $J=6$),
the existence of a magnetic ground state for an an isolated (i.e. assumed non-interacting)
Tb$^{3+}$ ion in Tb$_2$Ti$_2$O$_7$ is not  guaranteed by Kramers' theorem \cite{Gingras-TbTO-PRB-2000}.
To investigate whether Tb$^{3+}$ is magnetic, 
the so-called single-ion crystal field (CF) problem must first be
solved \cite{Gingras-TbTO-PRB-2000}.
 For example, for a perfect cubic ionic environment, theory predicts that Tb$^{3+}$ would
either have a singlet or non-magnetic doublet single-ion CF ground state \cite{Lea}.
However, in Tb$_2$Ti$_2$O$_7$, with its Fd$\bar 3$m symmetry, the
Tb$^{3+}$ environment displays a very large trigonal distortion away from cubic symmetry \cite{Gingras-TbTO-PRB-2000}.
Early investigations found that this distortion endows Tb$^{3+}$ with a magnetic CF doublet
characterized by  two mutually time-reversed conjugate states, 
$\vert \psi_0^+\rangle $ and $\vert \psi_0^-\rangle $ \cite{Gingras-TbTO-PRB-2000}.
The states, $\vert \psi_0^\pm \rangle$,  are such that all matrix elements of the raising and lowering
angular momentum operator, $J^\pm$, vanish while
$\langle \psi_0^\pm \vert J^z \vert \psi_0^\pm \rangle  = \pm \vert \langle J^z \rangle \vert
\approx 3.4$ \cite{Gingras-TbTO-PRB-2000}.
Since $\langle \psi_0^\pm \vert J^\mu \vert \psi_0^\pm \rangle  
= \pm \vert \langle J^z\rangle \vert \delta_{\mu,z}$,
the Tb$^{3+}$ moment within its CF ground state can be described by a 
classical Ising spin with a moment that points ``in'' or ``out'' of the reference  primitive tetrahedral unit
cell to which it belongs.
This makes Tb$_2$Ti$_2$O$_7$ a  relative of the Ising spin ice compounds 
\cite{Gingras-TbTO-PRB-2000,TTOHeff}.

The lowest excited CF state is
also a doublet, $\vert \psi_{\rm e}^{\pm} \rangle$, 
 at an energy approximately 1.6 meV $\sim$ 18 K above the ground doublet 
\cite{Gingras-TbTO-PRB-2000,Gardner-TbTO-PRB-2001,Mirebeau-TbXO-2007}.
Recent theoretical work \cite{TTOHeff}
has argued that the proximity of Tb$_2$Ti$_2$O$_7$ to the zero temperature transition from
an ``all-in/all-out'' ${\bf q}=0$ N\'eel to a spin ice state \cite{Hertog-PRL-2000}, 
along with the exchange and dipole-dipole interaction-induced admixing of
$\vert \psi_0      ^\pm \rangle$ and
$\vert \psi_{\rm e}^\pm \rangle$, constitute two key ingredients 
as to why Tb$_2$Ti$_2$O$_7$ does not develop long range order.

In conventional (unfrustrated) magnets, spin-lattice couplings typically
play an insignificant role in the development of long range order.
In highly frustrated magnetic systems, however, spin-lattice couplings
can lead to a combined magnetic-lattice  (``spin-Peierls'') 
transition to long range magnetic order
that reduces the magnetic frustration \cite{Villain-ZPhysB}, 
as observed in the highly frustrated antiferromagnet ZnCr$_2$O$_4$ spinel 
compound \cite{Lee-ZnCr2O4}. 
In Tb$_2$Ti$_2$O$_7$, Mamsurova and co-workers long ago reported 
an unusually large  anomalous field-dependent thermal expansion, indicating
an important spin-lattice coupling in this material \cite{Mamsurova}.
In more recent x-ray diffraction experiments, 
Ruff {\it et al.} found evidence for a tendency of the 
Tb$_2$Ti$_2$O$_7$ lattice to undergo a cubic tetragonal deformation but, 
down to 300 mK and in zero field, no  equilibrium cubic to 
tetragonal transition was observed \cite{Ruff-tetragonal}.   
Only in the presence of 
very high magnetic fields ($\sim 29$ T), does the system show
any evidence for such a structural 
phase transition \cite{Ruff-highfield}.
In contrast,  Chapuis and co-workers  very recently argued that 
a tetragonal deformation does actually exist in Tb$_2$Ti$_2$O$_7$ in zero
field that splits
the above  $\vert \psi_0^\pm \rangle$ doublet into two
{\it  non-magnetic singlets}
 separated by an energy scale, $\delta$, with  
$\delta \approx 1.8$ K \cite{Chapuis}.
They invoke previously published
inelastic neutron scattering (INS) data \cite{Mirebeau-TbXO-2007}
to suggest  that such a large splitting 
of the ground doublet compared to the exchange and dipolar interactions
between Tb$^{3+}$ ions is responsible for inhibiting the spontaneous development
of long range order.  
Reference ~[\onlinecite{Chapuis}] 
also present data for the temperature dependence of the magnetic entropy, $S(T)$,
which, they claim, by falling below R$\ln 2$ at low temperature,
further supports the evidence for a split doublet.
Very recently, Bonville and collaborators \cite{Bonville-singlet}
have built further
on Chapuis {\it et al.}'s split doublet picture to advocate
that the failure of Tb$_2$Ti$_2$O$_7$ to order is due to the sub-critical value
of the  interactions compared to the singlet-singlet gap $\delta$.

In this paper, we argue that the evidence for a split doublet of
energy scale as large as $\delta \sim 1.8$ K in Tb$_2$Ti$_2$O$_7$  as proposed
 in Refs.~[\onlinecite{Chapuis,Bonville-singlet}] is not compelling.
Using new high resolution INS data, we show that there is no evidence
for a split doublet in this material with an energy splitting greater than 0.2 K, a factor
10 or so {\it smaller} than the proposed~\cite{Chapuis,Bonville-singlet}
singlet-singlet gap $\delta$.
Secondly, we argue that by neglecting correlations that develop in the
collective paramagnetic (spin liquid) phase of Tb$_2$Ti$_2$O$_7$, 
the authors of Ref.~[\onlinecite{Chapuis}] are in principle
unable to draw any conclusion about the nature of the CF state of Tb$^{3+}$ in 
Tb$_2$Ti$_2$O$_7$ on the basis of a measurement of $S(T)$.
We illustrate that point via the calculation of $S(T)$ for a toy model which,
while possessing a ground state doublet and lacking a transition to long range order, does
display an $S(T)$ that falls {\it below} R$\ln 2$ at low temperature.


{\it Inelastic neutron scattering results} $-$
We first proceed to show that a singlet-singlet gap $\delta \sim 1.8$ K is 
inconsistent with high energy-resolution  inelastic neutron scattering
data and that, in fact, Tb$_2$Ti$_2$O$_7$ displays quasielastic 
magnetic spectral 
weight down to energies of at least 0.02 meV, approximately an order of magnitude 
lower in energy than the value $\delta \sim 1.8$ K 
reported in Refs.~[\onlinecite{Chapuis,Bonville-singlet}].

Time-of-flight neutron scattering data was obtained on single crystal 
Tb$_2$Ti$_2$O$_7$
in zero and finite magnetic field applied along the $[1 {\bar 1} 0]$ (vertical) direction at T=0.4 K.  These measurements used 
the Disk Chopper Spectrometer (DCS) 
at NIST  and employed incident neutrons of wavelength $\lambda$=4.8 $\AA$ resulting in an energy resolution of 0.1 meV. 
 The $H=0$, zero magnetic field data set, after integration along the [HH0] direction, perpendicular to the (00L) 
direction and within the horizontal scattering plane, is plotted in Fig. 1 a).
The quasi-elastic magnetic scattering peaks at {\bf Q}=002 (Q=1.2 $\AA^{-1}$).
Previous work ~\cite{Gardner-TbTO-PRB-2003,Gardner-TbTO-PRB-2001} has shown that 
the corresponding
 diffuse scattering is distributed in reciprocal space into a well known checkerboard pattern.
Figure 1 b) shows cuts in energy of  this zero magnetic field data set 
and also of a data set with a $H=3$ T magnetic field applied along the $[1 {\bar 1} 0]$  direction. 
 These cuts simulate constant-{\bf Q} scans, 
although they integrate the scattering
data in the [HH0] direction (shown in Fig.1 a) 
 and also in the [00L] direction between L=1.6 and 1.8.  
Two features are noteworthy: while there is a shoulder to this particular cut 
of the zero field quasi-elastic scattering near $E\sim$ 0.2 meV, 
there is a continuous and monotonically increasing distribution of magnetic 
spectral weight as the energy decreases down to zero energy.  
The nature of this shoulder near 0.2 meV is subtle
 for the zero magnetic field data set.
Indeed, low energy quasi-elastic magnetic
 scattering  is obscured by relatively strong nuclear 
incoherent elastic scattering which dominates the elastic signal within 
the 0.1 meV energy resolution of this measurement.  However, this
 nuclear incoherent contribution to the elastic scattering can be 
estimated and removed by examination of the $H=3$ T data set in Fig. 1 b).
Field-induced long range order  \cite{Rule-TbTO-PRL-2006}
 leads to a splitting-off 
of the quasi-elastic magnetic scattering from the resolution-limited nuclear 
incoherent elastic scattering.
Thus the strength of the nuclear incoherent contribution can be determined.  
Consider the two, otherwise identical cuts shown in Fig. 1 b), both at T=0.4 K, and  
taken at $H=0$  (top) and at $H=3$ T (bottom). 
One observes scattering at $E=0$ with an intensity of $\sim$ 9 for $H=0$ where 
the zero energy scattering has contributions from both nuclear
 incoherent elastic scattering and quasi-elastic magnetic scattering.  
In the bottom of  panel b), for $H=3$ T, an $E=0$ intensity 
of $\sim$ 3.5 is measured, which has contributions from nuclear incoherent scattering alone.
Hence the intensity of the elastic magnetic scattering at $H=0$ is $\sim$ 5.5 in the intensity units
 employed in Figs 1 b); at least a factor of two higher 
than the intensity associated with the shoulder near $\sim$ 0.2 meV.  
Therefore, the distribution of magnetic scattering intensity does 
indeed peak at zero energy and extends out to $\sim 0.3$  meV at low
 temperatures in $H=0$. In other words, there is no obvious 
mode centred at an energy of 1.8 K (0.16 meV) in the $H=0$ data

\begin{figure}[h]
\includegraphics{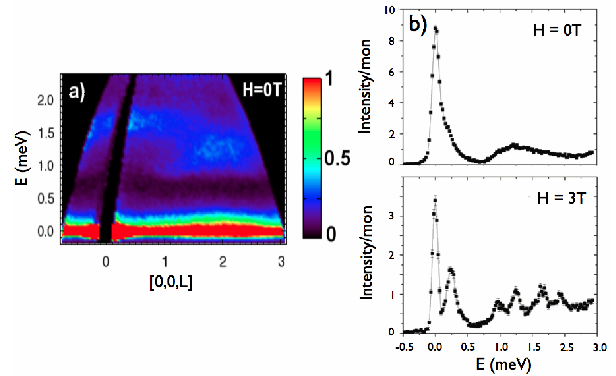}
\caption{\label{} a) High energy-resolution inelastic neutron scattering data taken with 
$\lambda$=4.8 $\AA$ neutrons on DCS \cite{Rule-TbTO-PRL-2006}. b) Energy cuts of the data shown 
in a) as described in the text.  Quasi-elastic magnetic scattering at T=0.4 K and zero
 applied magnetic field peaks at zero energy.  The crystal field excitation in 
zero field (top right panel) is anomalously broad in energy, compared with
 resolution-limited spin waves seen in the bottom right panel within the field-induced ordered phase.}
\end{figure}

To more definitely assess whether a finite energy excitation may
exist at $E\sim \delta \sim 0.16$ meV $\sim 0.18$ K, 
very high energy-resolution neutron scattering was 
carried out on the same single crystal of 
Tb$_2$Ti$_2$O$_7$, in zero magnetic field, as well with several different magnetic field strengths 
applied along the $[ 1 {\bar 1} 0]$ direction.  This data, taken with $\lambda=9$ $\AA$ incident
 neutrons on DCS, is shown in Fig. 2.  The scattering data shown in Fig. 2 integrates 
the raw inelastic scattering data
over all $\bf Q$ surveyed, which extends out in reciprocal space
 to the centre of the aforementioned diffuse magnetic scattering checkerboard
 at ${\bf Q}=002$  ~\cite{Gardner-TbTO-PRB-2003,Gardner-TbTO-PRB-2001}. 
 This very high energy-resolution data 
clearly shows the quasi-elastic magnetic scattering to increase {\it continuously} 
with decreasing energy down to the energy resolution of the measurement,
 $\sim$ 0.02 meV $\sim$ 0.2 K, an order of magnitude lower than 
the singlet-singlet gap $\delta\sim 1.8$ K invoked in Refs.~[\onlinecite{Chapuis,Bonville-singlet}].

\begin{figure}[h]
\includegraphics{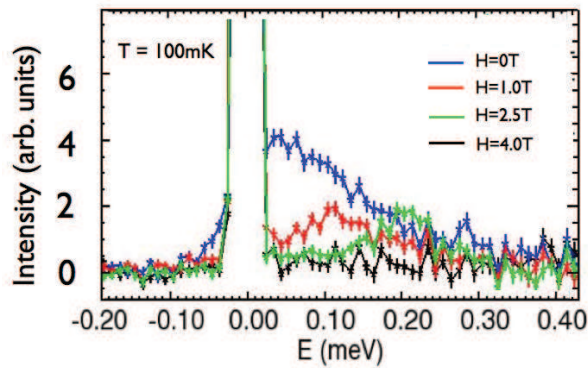}
\caption{\label{}Very high energy-resolution inelastic neutron scattering data 
employing $\lambda=9$ $\AA$ incident neutrons with DCS, as described in the text. 
 The zero field measurement shows the quasielastic magnetic distrution of scattering
 to extend down to at least 0.02 meV.}
\end{figure}

Figure 2 also shows that application of a sufficiently strong $[1 {\bar 1} 0]$ magnetic field  
($H \sim 4$  T) moves the quasi-elastic magnetic scattering out of the field of view 
of the figure, revealing a low  background which is energy-independent. 
 As a function of increasing applied magnetic field, the quasi-elastic 
scattering is strongly depleted at low energies, and an inelastic feature 
is observed for field strengths above $H \sim 1$ T.  This excitation moves to 
higher energies with increasing field strength.  However, 
and {\it crucially significant} for our argument,
 we see that even  with an applied field of $H=1$ T (red curve),
 the broad inelastic peak is at an energy of $\sim$ 0.1 meV $\sim 1$ K, 
roughly a factor of 2 lower in energy than the gap $\delta\sim 1.8$ K
 invoked   in Refs.~[\onlinecite{Chapuis,Bonville-singlet}] in zero field.

One could speculate that the quasielastic distribution of magnetic scattering in zero field 
out to $\sim$ 0.3 meV is a dispersive singlet-singlet excitation with 
a band width of twice the mean separation between the two states, such that its 
density-of-states fills in th quasi-elastic energy range. 
However, the intensity would then be the largest for the top of this band, 
as the density-of-states would be high where the dispersion is flat.
This is not observed --
the quasi-elastic scattering (Fig. 2) decreases monotonically with 
increasing energy.

One could also speculate a static and random Jahn-Teller distortion in which a
 broad distribution of singlet-singlet gaps is present in the spin liquid state.
This could lead to a distribution of gaps and induce monotonically-decreasing 
(as a function of energy) quasielastic scattering in zero field as seen in 
Figs. 1 and 2.
However, with decreasing temperature, these static gaps would get progressively frozen out, 
resulting in a larger and larger fraction of the system being in a non-magnetic state. 
 The $1/T_1$ muon spin relaxation rate is large and flat in temperature from $\sim$ 
1 K down to 0.05 K \cite{Gardner-TbTO-PRL-1999}. This would seem
to rule  out this scenario since, at 0.05 K,  
 excited singlets 
with an energy gap larger than 0.005 meV would be frozen out.
Similarly, neutron spin echo (NSE) results in 
Ref.~[\onlinecite{Gardner-TbTO-PRB-2003}]  rules out a large 1.8 K gap in the system. 
NSE with sub-$\mu$eV resolution and polarised neutron diffraction 
\cite{Gardner-TbTO-PRB-2003} both show an increase 
in the magnetic scattering below 300 mK suggesting a magnetic ground state or an extremely 
small gap that is thermally active at 50 mK.

Taken altogether, we conclude there is no compelling evidence for a 
well-defined singlet-singlet gap in Tb$_2$Ti$_2$O$_7$
 in zero field at low temperatures. 
 Its quasi-elastic, magnetic spectrum is not substantially different from 
that displayed by Ho$_2$Ti$_2$O$_7$
 \cite {Clancy-HoTO-PRB} just above its frozen spin ice ground state, 
albeit with a higher energy scale.


{\it Residual entropy} $-$ 
Having established that there is no direct evidence for a split doublet with
an energy gap larger than 0.02 meV $\sim$ 0.2 K in Tb$_2$Ti$_2$O$_7$, we now
 address the interpretation given in Ref.~[\onlinecite{Chapuis}] of
 low temperature magnetic entropy
data determined from the heat capacities of \tso\ and Tb$_2$Ti$_2$O$_7$.
 The magnetic entropy results
 for \tso\ are given in Fig. 2 of Ref.~[\onlinecite{Chapuis}]. In this data, 
the nuclear and phonon contributions have been subtracted.
One sees that roughly 
R$\ln 4$ is lost upon cooling the sample from $20$ K down to $40$ mK.
 This  is consistent with having no
extensive residual entropy and with magnetic ions having four states lying beneath
$20$ K. 
The data for Tb$_2$Ti$_2$O$_7$ 
also indicate a similar loss of magnetic entropy of R$\ln 4$  
below $20$ K \cite{Chapuis}. 
The authors of Ref.~[\onlinecite{Chapuis}] proceed to make the following argument: 
Suppose that there is a doublet-doublet crystal field
scheme with a gap $\Delta$. Then, at high temperatures exceeding the scale of 
the gap, the entropy would be R$\ln 4$ whereas at low
temperatures, $T\ll \Delta$, the entropy would saturate at R$\ln 2$. 
Therefore, in this scheme, the total entropy variation should
be R$\ln 2$. Since the entropy variation is observed to be R$\ln 4$
 in both \tso\ and Tb$_2$Ti$_2$O$_7$,  Ref.~[\onlinecite{Chapuis}] concludes
that the doublet-doublet energy level scheme must be incorrect. 

In an attempt to  
support their argument, the authors present a calculation of the entropy for a crystal field scheme
consisting of a pair of low-lying singlets separated by a tuneable energy gap 
of $\delta$ and one excited doublet at energy
$\Delta>\delta$ above the ground singlet. Upon increasing $\delta$ from zero, 
the lowest temperature entropy drops to zero for $T\lesssim
\delta$ and exhibits a plateau at  R$\ln 2$ for
 $\delta \lesssim T \lesssim \Delta$
(see Fig. 3 of Ref.~[\onlinecite{Chapuis}]).
 The width of the plateau decreases as $T$ increases  and for
$\delta\approx 1.8$ K, the entropy variation for this model roughly
 matches that for \tto\ (the fit is shown in Fig. $4$ of
Ref.~[\onlinecite{Chapuis}]). This is presented as evidence for a 
singlet-singlet crystal field scheme in \tto. A similar conclusion
is reached for \tso.

However, the authors of 
Ref.~[\onlinecite{Chapuis}]
have not ruled out the possibility that the magnetic entropy 
is lost through the effects of interactions and the concomitant build-up of correlations as the 
temperature is decreased below 20 K. This is particularly evident in 
 Tb$_2$Sn$_2$O$_7$ which exhibits
 a phase transition at 0.87 K to a long range ordered phase. This implies that the residual 
entropy for $T < 0.87$ K K should have no extensive contribution, 
 as appears to be borne out by the analysis of the specific heat capacity data in  Ref.~[\onlinecite{Chapuis}].
It follows that one must account for the details of the transition in order to extract
 information about the single ion level structure from entropy data obtained from the
 magnetic heat capacity while no attempt was made to carry out such an analysis in Ref.~[\onlinecite{Chapuis}]

\begin{figure}[h!]
\begin{center}
\subfigure
{
\includegraphics[width=0.48\columnwidth,clip]{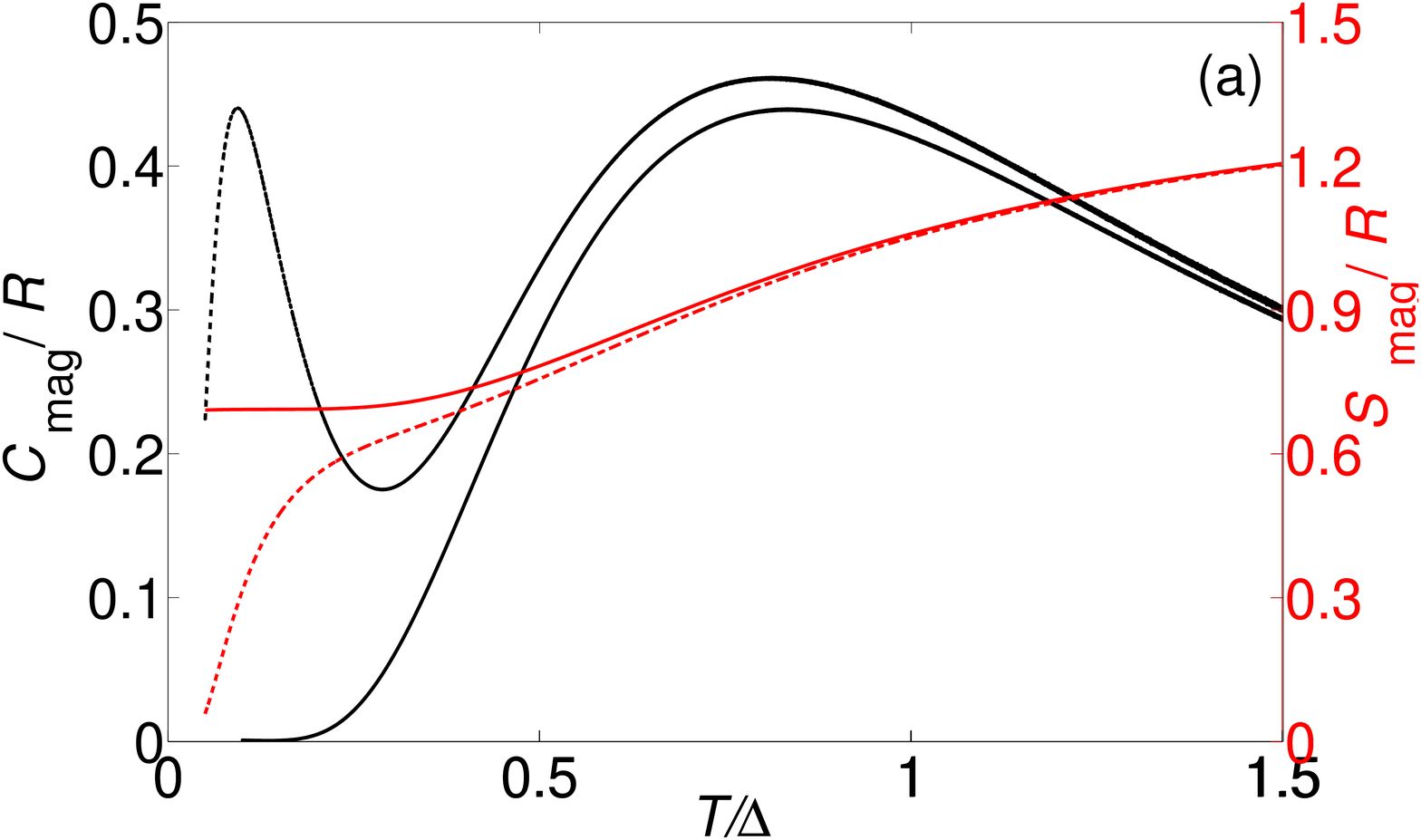}}
\subfigure
{
\includegraphics[width=0.48\columnwidth,clip]{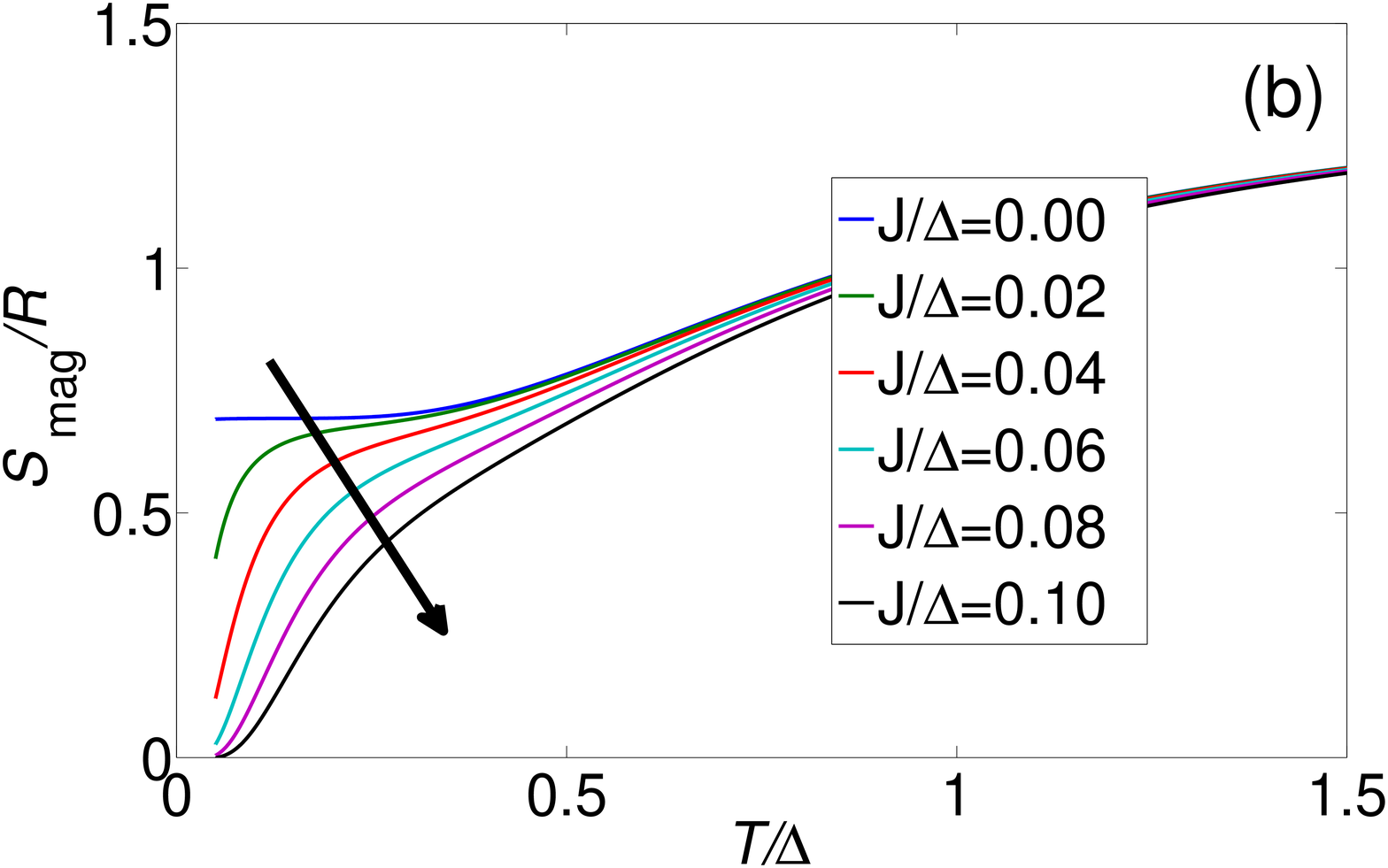}}
\caption{\label{fig:qm/complexfunctions}
Panel (a) shows the heat capacity, $C_{\rm mag}$ (black)
and entropy, $S_{\rm mag}$ (red),
for a one-dimensional spin $3/2$ Ising model with
$J=0$ (solid lines)
and $J/\Delta = 0.05$ (dashed lines).
The plateau in the entropy at low
temperatures is at $\ln 2$.
Panel (b) shows the entropy for a one-dimensional
spin $3/2$ Ising
 model with $J/\Delta$ from $0$ to $0.1$.
 The plateau gradually becomes smaller as $J/\Delta$ increases.
}
\end{center}
\end{figure}

We now turn to the case of \tto, which does not
exhibit a transition to long range order down to the lowest observed temperature of
$50$ mK \cite{Gardner-TbTO-PRL-1999,Gardner-TbTO-PRB-2003}.
Nevertheless, interactions are significant in this material as can be inferred 
from the Curie-Weiss temperature \cite{Gingras-TbTO-PRB-2000}, the
paramagnetic diffuse scattering \cite{Gardner-TbTO-PRL-1999,Gingras-TbTO-PRB-2000,Kao,TTOHeff} 
and the dispersion on inelastic neutron scattering peaks 
for example \cite{Gardner-TbTO-PRB-2003,Kao,Mirebeau-TbXO-2007}. 
Indeed, the  overall scale of exchange
interactions in \tso\ and \tto\ 
have been estimated to be of similar magnitude  \cite{Mirebeau-TbXO-2007}.
Although the interactions  in \tto\ do not manifest themselves as a
phase transition, they potentially have a large effect on the entropy as we now
 illustrate with a toy model.
 We consider a spin-$3/2$ Ising model on a chain
with ferromagnetic nearest neighbor exchange $J$ ($J>0$), 
\[ 
H = \sum_{i} \left( -\Delta (S_{i}^{z})^{2} - J S_{i}^{z}S_{i+1}^{z} \right).
 \]
Of importance for our argumentation, this model has 
(i) a ground state doublet, (ii) no phase transition to long range order at
nonzero temperature and (iii) a paramagnetic entropy of R$\ln 4$ similarly
to Tb$_2$Ti$_2$O$_7$.
This model can be solved exactly by splitting the partition function trace 
into a product of transfer matrices. Consider first the case $J=0$.
The specific heat $C_{\rm mag}$  shows a single peak (Schottky anomaly)
corresponding to the loss of entropy, $S_{\rm mag}$, 
as the excited $S^z=\pm 3/2$ 
levels are depopulated upon cooling [solid lines in top panel of Fig. 3(a)]. 
The entropy exhibits two plateaus -- one at low
temperature (R$\ln 2$) and another at high temperature (R$\ln 4$). 
This is the analogue of the non-interacting 
doublet-doublet model considered in Ref.~[\onlinecite{Chapuis}].

Switching on the exchange $J$ causes FM correlations to build
up. There is no transition, but there is a peak in $C_{\rm mag}$ 
beneath the Schottky anomaly at $T/J \sim 0.1$. 
This is reflected in the magnetic entropy, $S_{\rm mag}$, dropping below the R$\ln 2$ value. 
The R$\ln 2$ plateau gradually shrinks as $J$ increases [Fig. 3(b)].
We conclude that the disappearance of an R$\ln 2$ entropy plateau can occur 
when interactions are introduced {\it even
in the absence of a phase transition} and, therefore, 
cannot be attributed definitively to a splitting of the doublet
ground state without interactions.


{\it Conclusion} $-$
We have reconsidered the scenario advanced in Refs.~[\onlinecite{Chapuis,Bonville-singlet}]
 whereby long range order for Tb$_2$Ti$_2$O$_7$
is evaded because the single ion crystal field states are split in zero
field,  ostensibly due to a tetragonal distortion breaking the Fd$\bar{3}$m
symmetry. We have presented high energy resolution neutron scattering data to
re-examine the case for a singlet-singlet splitting, finding no
evidence for excitations that would indicate a splitting greater than $0.2$ K. 
We have also argued that
measurements determining the residual entropy cannot be used to draw
conclusions about the single ion spectrum without considering the build-up of short-range
correlations at low temperatures. In conclusion, the nature of the low
temperature state of Tb$_2$Ti$_2$O$_7$ remains a remarkable and unsolved
problem in the field of frustrated magnetism.
One may anticipate further progress in light 
of the constraints that we and others are placing on possible
scenarios to explain the spin liquid behavior in this material.   

This research was funded by the NSERC of Canada and the Canada
Research Chair program (M. G., Tier I).

\end{document}